\documentclass[aps,prl,reprint,superscriptaddress,footinbib]{revtex4-1}
\usepackage{amsmath,bm}
\usepackage{graphicx}
\usepackage{multirow}
\usepackage{hyperref}
\usepackage{comment, color, xcolor}
\usepackage{ulem}

\newcommand{\be}{\begin{equation}}
\newcommand{\ee}{\end{equation}}
\newcommand{\bes}{\begin{subequations} \begin{align} }
\newcommand{\ees}{\end{subequations}\end{align} }
\newcommand{\bea}{\begin{eqnarray}}
\newcommand{\eea}{\end{eqnarray}}

\newcommand{\eq}[1]{Eq.~\eqref{eq:#1}}

\newcommand{\fig}[1]{Fig.~\ref{fig:#1}}

\newcommand{\tab}[1]{Table~\ref{tab:#1}}

\newcommand{\yl}[1]{{\color{blue} #1}}

\begin{document}

\title{NRQCD Re-Confronts LHCb Data on Quarkonium Production within Jets}

\author{Yunlu Wang}
\email{yunluwang20@fudan.edu.cn}
\affiliation{Key Laboratory of Nuclear Physics and Ion-beam Application (MOE) and Institute of Modern Physics, Fudan University, Shanghai, China 200433}

\author{Daekyoung Kang}
\email{kang@fudan.edu.cn}
\affiliation{Key Laboratory of Nuclear Physics and Ion-beam Application (MOE) and Institute of Modern Physics,
Fudan University, Shanghai, China 200433}
\affiliation{Department of Physics, Korea University, Seoul 02841, Korea}

\author{Hee Sok Chung}
\email{heesokchung@gwnu.ac.kr}
\affiliation{Department of Mathematics and Physics,
Gangneung-Wonju National University, Gangneung 25457, Korea}
\affiliation{Department of Physics, Korea University, Seoul 02841, Korea}

\date{\today}

\begin{abstract}
We compare LHCb measurements of $J/\psi$ and $\psi(2S)$ transverse momentum
distributions within jets with QCD calculations, which may be crucial in
understanding the quarkonium production mechanism. 
Our theoretical calculations are based on the fragmenting jet function
formalism, while the nonperturbative formation of quarkonia is described 
by the nonrelativistic QCD factorization formalism. 
We include the newest refinements in the perturbative calculation 
including resummation of threshold and DGLAP logarithms. 
We find that the $\psi(2S)$ data has the potential to discriminate between 
the different production mechanisms proposed in the literature.
\end{abstract}

\maketitle
{\it Introduction---}Understanding the heavy quarkonium production mechanism is 
one of the most significant challenges in Quantum Chromodynamics (QCD). 
Existing approaches based on nonrelativistic QCD (NRQCD) descriptions 
of inclusive production rates as functions of the transverse momentum $p_T$ 
had difficulty in pinpointing the exact production
mechanism~\cite{Chung:2018lyq, Chung:2022uih}. 
This happens because inclusive cross sections generally do not have enough
degrees of freedom to determine a sufficient number of NRQCD long-distance
matrix elements (LDMEs), which govern the nonperturbative formation of the
quarkonium. Because of this, a more differential observable such as the
transverse momentum distribution of the quarkonium within a jet has been
proposed a way to scrutinize the production mechanism~\cite{Baumgart:2014upa, 
Bain:2016rrv, Kang:2017yde}.

The momentum distribution of $J/\psi$ within a jet has been measured as
functions of the transverse momentum fraction $z$ by LHCb~\cite{LHCb:2017llq}, 
CMS~\cite{CMS:2021puf}, and STAR~\cite{Yang:2021smr} experiments, and recently 
LHCb reported measurements for $\psi(2S)$ as well~\cite{LHCb:2024ybz}. 
An initial theoretical study~\cite{Bain:2017wvk} based on the fragmenting jet
function (FJF) formalism~\cite{Baumgart:2014upa} offered comparisons of
predictions from different LDME sets to the LHCb $J/\psi$ data, demonstrating
the advantage of this novel approach. However, this has not led to a
definitive determination of the NRQCD LDMEs and the quarkonium production 
formalism, because the authors of Ref.~\cite{Bain:2017wvk} only considered a
limited class of LDME sets, and have not considered important logarithmic
corrections near the kinematical threshold, which was unknown at that time. 
Also, the calculation in Ref.~\cite{Bain:2017wvk} did not match
exactly the kinematical conditions of the LHCb experiment. 
Therefore, an updated analysis of both $J/\psi$ and $\psi(2S)$ data involving
more extensive LDME sets with more refined perturbative calculations is much
desired.

In this letter, we provide improved theoretical predictions for the
transverse-momentum distribution of $J/\psi$ and $\psi(2S)$ produced in jets.
We employ a FJF formalism developed in Ref.~\cite{Kang:2016ehg} that can
facilitate the same kinematical conditions of LHCb measurements. In the
perturbative calculation of the FJF we resum not only the DGLAP logarithms but
also the threshold double logarithms, which is crucial in obtaining a
nonsingular, positive definite distributions near the threshold
$z=1$~\cite{Chung:2024jfk}. 
For the LDMEs, we choose three representative examples in
Refs.~\cite{Brambilla:2022ayc, Bodwin:2015iua, Butenschoen:2011yh}, which
effectively span the ranges of available LDME sets in the literature. 
We then compare our results with LHCb data~\cite{LHCb:2017llq, LHCb:2024ybz}
to assess the power of the quarkonium-in-jet measurements in resolving the
quarkonium production mechanism.

{\it NRQCD factorization and LDMEs---}
In the NRQCD factorization formalism, the inclusive production rate of a heavy
quarkonium $H$ is given by 
\be\label{eq:sigmaH}
\sigma_H
 = \sum_n
\sigma_{ Q\bar Q (n)}\,
\langle \mathcal{O}^{H}(n) \rangle\,,
\ee
where $\sigma_{ Q\bar Q (n)}$ is the production rate of a heavy quark $Q$ and 
antiquark $\bar Q$ pair in the state $n$, 
and $\langle \mathcal{O}^{H}(n) \rangle$ is the LDME that describes the
evolution of the $Q \bar Q(n)$ into the meson $H$. 
We use the spectroscopic notation $^{2S+1}L_J^{[c]}$ to denote the state $n$,
where $S$ and $L$ represent the spin and orbital angular momentum,
respectively, and $c= 1$, 8 denote the color. 
In practice, the sum in Eq.~(\ref{eq:sigmaH}) is truncated to include only the
most important contributions. For production of $J/\psi$ or $\psi(2S)$, they
include $n={}^{3}S_1^{[1]}$, ${}^{3}S_1^{[8]}$, ${}^{1}S_0^{[8]}$, and 
${}^{3}P_J^{[8]}$. 
While the $\sigma_{ Q\bar Q (n)}$ are perturbatively calculable, it is in
general not known how to compute the color-octet LDMEs from first principles. 
Hence, the three color-octet LDMEs have been determined from fits to production
data. This approach has not been successful in obtaining a satisfactory
description of the $J/\psi$ or $\psi(2S)$ production mechanism, as the LDME sets 
obtained in this way can vary significantly depending on the choice of data and
the accuracy of the perturbative calculation~\cite{Chung:2018lyq,
Chung:2022uih}. 

\begin{widetext}

\begin{table}[thb]
    \centering
    \begin{tabular}{cccccc}
        \hline
        \multirow{2}{*}{ \phantom{Empty}} &\multirow{2}{*}{\phantom{Empty}} 
        & $ \langle \mathcal{O}(^{3}S^{[1]}_{1}) \rangle$ 
        & $ \langle \mathcal{O}(^{3}S^{[8]}_{1}) \rangle$ 
        & $\langle \mathcal{O}(^{1}S^{[8]}_{0}) \rangle$ 
        & $\langle \mathcal{O}(^{3}P^{[8]}_{0}) \rangle / m_c^2$ \\ 
       & &\phantom{Empty} $\times~\text{GeV}^3$ & \phantom{Empty}$\times~10^{-2}~\text{GeV}^3$ 
        & \phantom{Empty}$\times~10^{-2}~\text{GeV}^3$ & \phantom{Empty}$\times~10^{-2}~\text{GeV}^3$ \\ \hline\hline
        
\multirow{3}{*}{ $J/\psi$  }
& Brambilla et al.\cite{Brambilla:2022ayc}
      & $1.18 \pm 0.35$ & $1.40 \pm 0.42$ & $-0.63 \pm 3.22$ &$2.33 \pm 0.83$ \\ \cline{2-6} 

& Bodwin et al. \cite{Bodwin:2015iua}
  & $1.32 \pm 0.20$ & $-0.71\pm 0.36$ & $11.0 \pm 1.4$ & $-0.31 \pm 0.15$ \\ \cline{2-6}

& B\&K  \cite{Butenschoen:2011yh} 
    & $1.32 \pm 0.20$ & $0.22 \pm 0.06$ & $4.97 \pm 0.44$ &$-0.72 \pm 0.09$ \\ \hline\hline
 
\multirow{3}{*}{ $\psi(2S)$  }
& Brambilla et al. \cite{Brambilla:2022ayc}
     & $0.71 \pm 0.21$ & $0.84 \pm 0.25$ & $-0.37 \pm 1.92$ &$1.55 \pm 0.49$ \\ \cline{2-6}
& Bodwin et al. \cite{Bodwin:2015iua}
     & 0.76\cite{Eichten:1995ch} & $-0.16 \pm 0.28$ & $3.14 \pm 0.79$ & $-0.12 \pm 0.12$ \\  \cline{2-6}
& B\&K \cite{Butenschoen:2022qka}
     & 0.76\cite{Eichten:1995ch} & $0.054 \pm 0.003$ & $1.00 \pm 0.03$  &$-0.217\pm 0.005$\\ \cline{1-6}
        \end{tabular}
\caption{Three sets of LDMEs for $J/\psi$ and for $\psi(2S)$ in units of GeV$^3$ and $m_c=1.5$~GeV.} 
    \label{tab:ME}

\end{table}
 \end{widetext}

The LDME sets available in the literature can generally be classified into
three categories. 
\textit{Category 1}: Small ${}^1S_0^{[8]}$ LDME, with ${}^3S_1^{[8]}$ and ${}^3P_J^{[8]}$ having the same sign. The cross section is dominated by ${}^3S_1^{[8]} + {}^3P_J^{[8]}$. Examples include Refs.~\cite{Chung:2024jfk,Brambilla:2021abf,Han:2014jya,Zhang:2014ybe} and Ref.~\cite{Shao:2014yta} (minimized ${}^1S_0^{[8]}$).
\textit{Category 2}: Large ${}^1S_0^{[8]}$ LDME, with ${}^3S_1^{[8]}$ and ${}^3P_J^{[8]}$ having the same sign. The cross section is dominated by ${}^1S_0^{[8]}$. Examples include Refs.~\cite{Gong:2012ug,Bodwin:2015iua,Feng:2018ukp}, Table II (Fit D) of Ref.~\cite{Butenschoen:2022qka}, and Ref.~\cite{Shao:2014yta} (maximized ${}^1S_0^{[8]}$).
\textit{Category 3}: Small ${}^1S_0^{[8]}$, positive ${}^3S_1^{[8]}$, and negative ${}^3P_J^{[8]}$. Examples include Ref.~\cite{Butenschoen:2011yh} and Fits A \& B of Ref.~\cite{Butenschoen:2022qka}.
Predictions within each category would be similar. Representative sets for $J/\psi$ and $\psi(2S)$ are listed in \tab{ME}.

{\it Fragmenting jet function---}%
Given tensions between LDME sets, studying quarkonium distributions inside jets~\cite{Baumgart:2014upa,Bain:2016rrv,Kang:2017yde} provides a valuable probe. 
There are two variants of FJF: one defined from exclusive $n$-jet processes~\cite{Procura:2009vm,Liu:2010ng,Procura:2011aq,Jain:2011iu,Jain:2012uq,Bauer:2013bza,Ritzmann:2014mka} and the other from semi-inclusive jet processes~\cite{Kaufmann:2015hma,Kang:2016mcy,Kang:2016ehg,Dai:2016hzf,Dai:2017dpc}. 
We compute the cross section using semi-inclusive FJFs, which suit the LHCb experimental framework, in contrast to a prior analysis based on exclusive jets~\cite{Bain:2017wvk}. These FJFs depend on the jet's transverse momentum ($p_T$) and radius ($R$), the quarkonium's momentum fraction ($z_H = p_T^H/p_T^{\text{jet}}$), and the jet's momentum fraction relative to the initial parton ($z$).

When the jet virtuality $p_T R \gg m_H$, the FJF factorizes into a convolution of quarkonium fragmentation functions (FFs)~\cite{Braaten:1993rw,Braaten:1994vv,Braaten:1993mp} and Wilson coefficients, up to $\mathcal{O}(m_H^2/(p_T R)^2)$ corrections.
\be \label{eq:FJF}
\mathcal{G}_i^H\left(z, z_H, p_T R, \mu\right)=\sum_j  \mathcal{J}_{i j}\left(z, z_H, \mu\right) \otimes D_j^H\left(z_H, \mu\right),
\ee
where the coefficients $\mathcal{J}_{ij}$ describe splitting of a mother parton $i=\{q,g\}$ into a daughter parton $j$ that resides in the jet \cite{Kang:2016ehg}.
The FF $D^H_j$, which describes the probability of a parton $j$ fragmenting into a quarkonium $H$, is further factorized into short-distance coefficients and the LDMEs.
Note that the coefficients $\mathcal{J}_{i j}$ \cite{Kang:2016ehg} in \eq{FJF} are different from those for exclusive jet \cite{Baumgart:2014upa} in out-of-jet contributions.
The hierarchy $m_H \ll p_T R \ll p_T $ induces large logarithms, $\ln (p_T R /m_H)$ and $\ln R$, in the cross section, introducing significant theoretical uncertainties.
These uncertainties can be systematically reduced by resumming the logarithms through renormalization group evolution (RGE): 
from $m_H$ to $p_T R$ for FF and from $p_T R$ to $p_T$ for FJF.
The RGE for the FF is governed by the well-known DGLAP, while for the FJF, it follows DGLAP at the leading order and becomes DGLAP-like but different at higher orders \cite{Lee:2024tzc,Lee:2024icn}.
The FFs in $^3S_1^{[8]}$ and $^3P_J^{[8]}$ channels show divergent behavior in the threshold limit $z_H\rightarrow 1$.
For the first time in the FJF framework, we adopt the recently developed threshold resummation formula \cite{Chung:2024jfk},  which resums the double logarithms $\log(1-z_H)/(1-z_H)$ and resolves the catastrophic failure of FFs.

{\it Numerical calculation.---}
We perform the corresponding calculation
for the LHCb measurements, $J/\psi$ \cite{LHCb:2017llq} and  $\psi(2S)$ \cite{LHCb:2024ybz}.
The LHCb analysis uses anti-$k_t$ jets with \yl{a} radius parameter $R = 0.5$ selected within the pseudorapidity range $2.5 < \eta < 4.0$. 
For $J/\psi$ production, all events with jet transverse momentum $p_T > 20$ GeV are accumulated, generating a single momentum fraction $z_H$ distribution. The $\psi(2S)$ dataset provides finer resolution through binning in both jet-$p_T$ (7 bins spanning 8--60 GeV) and $\psi(2S)$-$p_T$ (6 bins spanning 5--40 GeV) and we focus on comparison to the three highest-$p_T$ bins in each category, where our predictions are accurate. 

Experimentally, quarkonia are reconstructed via their muon-pair decay channels.
LHCb's muon identification requires both decay products to satisfy pseudorapidity $2.0 < \eta < 4.5$ and transverse momentum $p_T > 0.5$ GeV, with additional momentum thresholds of $p_{\mu} > 5$ GeV for $J/\psi$ and $p_{\mu} > 6$ GeV for $\psi(2S)$. These cuts preferentially suppress low-$z_H$ events where the quarkonium carries a small fraction of the jet's momentum. Our calculations implement these selections by first modeling unpolarized quarkonium decays with isotropic $\mu^+\mu^-$ angular distributions in the quarkonium rest frame, then applying Lorentz boosts to the lab frame before enforcing the kinematic cuts on the boosted muons. This approach properly accounts for the $z_H$-dependent acceptance efficiency across the full kinematic range.

\onecolumngrid

\begin{figure}[tbh]
	\begin{center}
        \includegraphics[width=.3\textwidth]{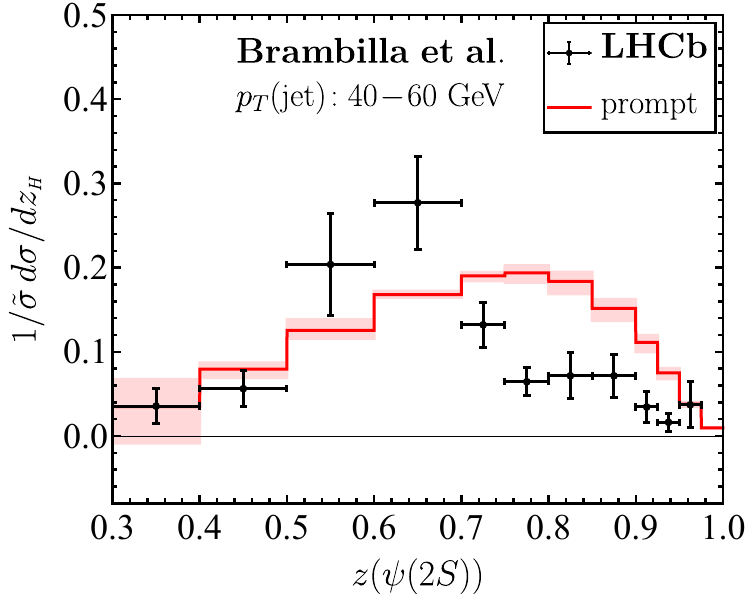}
	\includegraphics[width=.3\textwidth]{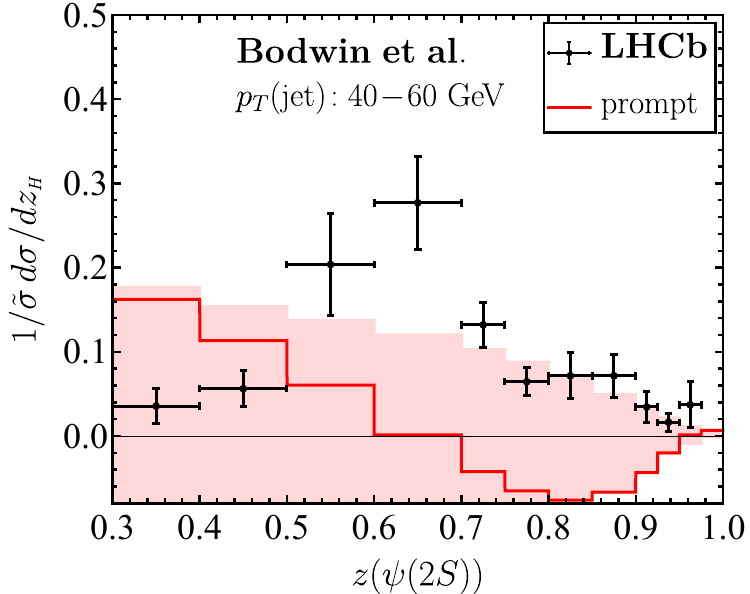}
        \includegraphics[width=.3\textwidth]{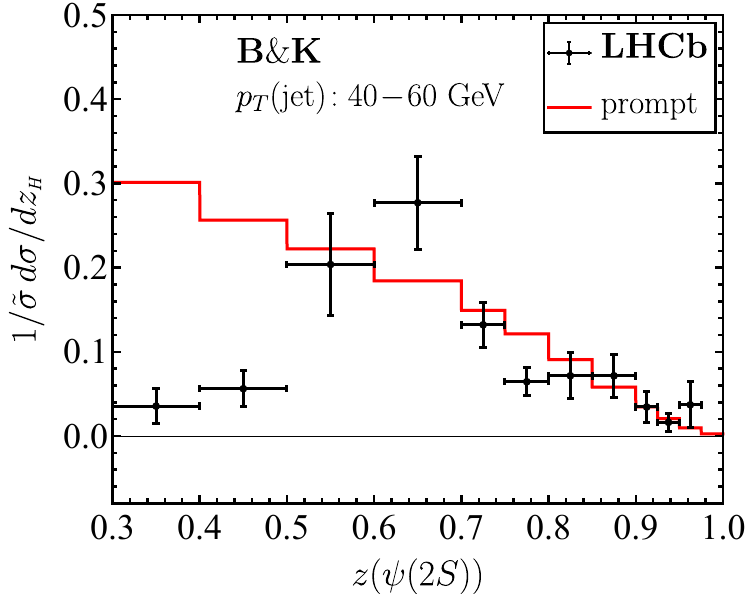}
        \includegraphics[width=.3\textwidth]{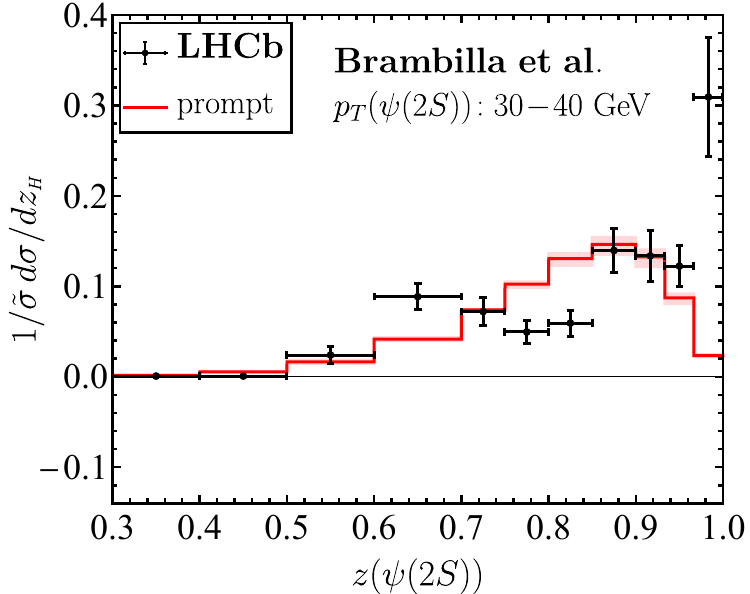}
	\includegraphics[width=.3\textwidth]{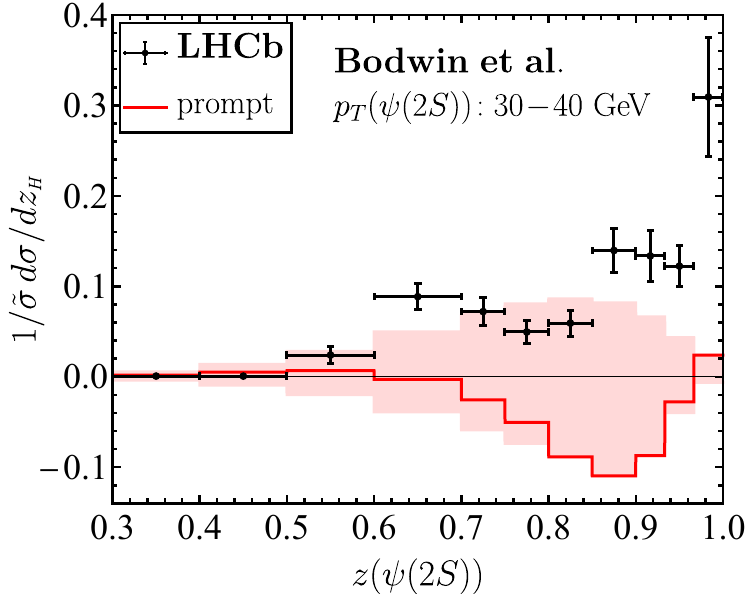}
        \includegraphics[width=.3\textwidth]{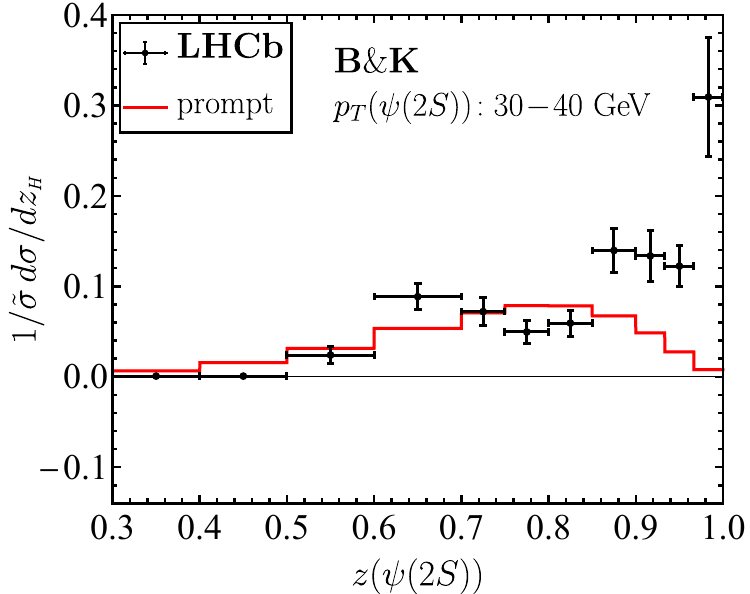}
	\end{center}\vspace{-4ex}
	\caption{Prediction for $z_{\psi(2S)}$ distributions for three sets of LDMEs in \tab{ME} and the LHCb measurements. The factor $\tilde{\sigma}$ follows LHCb's normalization.}
	\label{fig:psi2S_LHCb}
\end{figure}

\twocolumngrid
In our calculation, we convolute partonic cross sections at next-to-leading order (NLO)~\cite{Jager:2002xm} with FJFs for quark and gluon channels. The FJFs incorporate NLO coefficients $\mathcal{J}_{ij}$~\cite{Kang:2016ehg} and $\mathcal{O}(\alpha_s^2)$ FFs~\cite{Ma:2013yla} for the four channels listed in \tab{ME}. 
The NLO cross section expansion reaches $\mathcal{O}(\alpha_s^4)$, with the $^3S^{[8]}_1$ channel entering from $\mathcal{O}(\alpha_s^3)$ while the other three channels contribute at $\mathcal{O}(\alpha_s^4)$. 
We perform DGLAP evolution of threshold-resummed FFs from $2m_c$ to the jet scale $p_T R$  at leading-log (LL) accuracy, followed by FJF evolution from $p_T R$ to the hard scale $p_T$ at LL. 
This procedure achieves an accuracy of LL$R$ + LL${\text{threshold}}$ + NLO, representing the most refined perturbative calculation for this process.

{\it Prompt-$\psi(2S)$ results.---}%
\fig{psi2S_LHCb} displays our predicted $z_{\psi(2S)}$ distributions for the three LDME sets from \tab{ME}, comparing results in both a 40--60 GeV jet-$p_T$ bin (upper panel) and a 30--40 GeV $\psi(2S)$-$p_T$ bin (lower panel), with additional lower-$p_T$ predictions provided in the supplementary material.
The $z_{\psi(2S)}$ variation is performed through two complementary approaches: varying $\psi(2S)$-$p_T$ within a fixed jet-$p_T$ bin and varying jet-$p_T$ within a fixed $\psi(2S)$-$p_T$ bin.
 
While the LHCb data is normalized in $0 < z_H < 1$, our theoretical predictions are normalized to the corresponding area under the histogram within the mid-$z_H$ region ($0.5 < z_H < 0.8$), avoiding potential biases from the less reliable low-$z_H$ and high-$z_H$ regimes. 
The theoretical uncertainties are calculated using covariance matrices for all three LDME sets. As shown in \tab{ME}, the B\&K set exhibits significantly smaller uncertainties compared to the other sets.
For clarity in comparing LDME effects, we intentionally exclude perturbative uncertainties from scale variations. 

The three LDME sets produce distinctive predictions, yet none fully reproduces the LHCb data. The agreement appears to be region-dependent, with the Brambilla set providing a better description at low $z_H$ and the B\&K set performing better at high $z_H$, while a conclusion for the Bodwin set is difficult to draw owing to its large uncertainties.
All cases demonstrate dominant contributions from the $^3P_J^{[8]} + {}^3S_1^{[8]}$ channels, with $^3S_1^{[1]}$ being negligible and $^1S_0^{[8]}$ relatively minor compared to other octet contributions. Notably, the Bodwin set produces unphysical negative predictions across significant $z_H$ ranges due to its large negative $^3S_1^{[8]}$ contribution overwhelming the positive $^3P_J^{[8]}$ contribution.
In upper panel, a low $z_H$ region receives power corrections of an order of $(m_H/p_T^{H})^2$ that is not included in our results and shows significant deviations from the data, while in lower panel we observe better agreement with the data since $p_T^H$ is held fixed and the correction remains small in low $z_H$ region.
\onecolumngrid

\begin{figure}[htb]
	\begin{center}        
	\includegraphics[width=.3\textwidth]{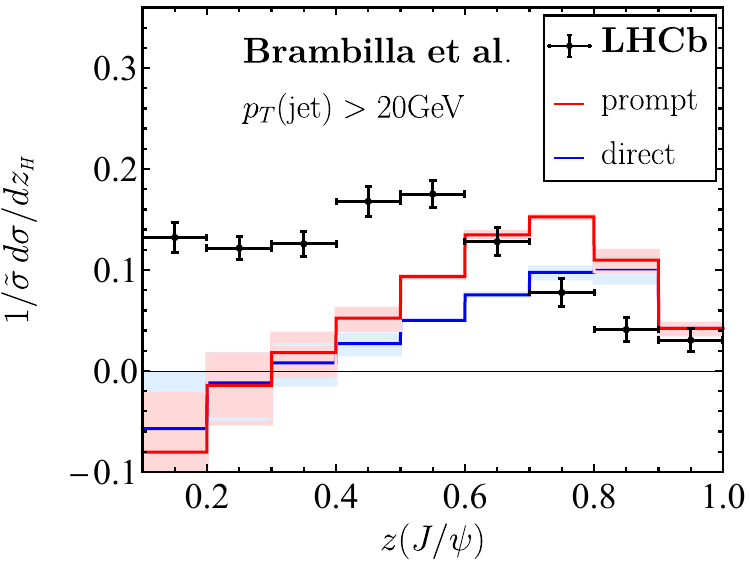}
	\includegraphics[width=.3\textwidth]{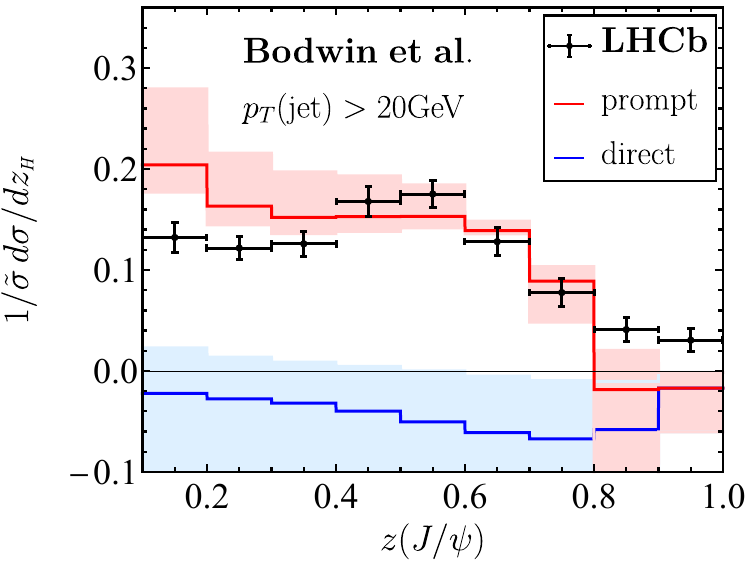}
        \includegraphics[width=.3\textwidth]{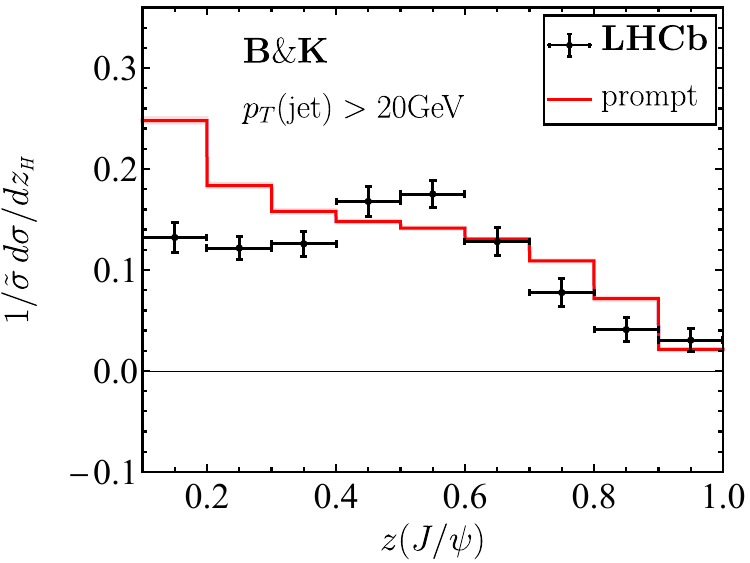}      
	\includegraphics[width=.3\textwidth]{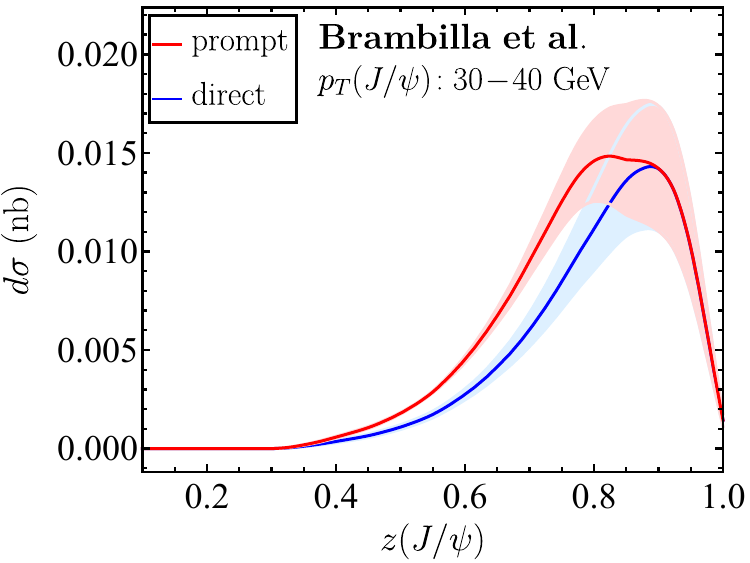}
	\includegraphics[width=.3\textwidth]{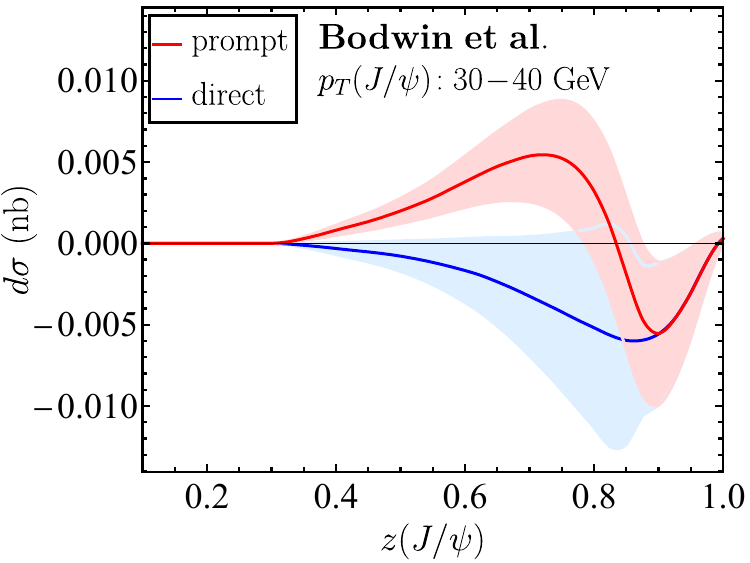}
        \includegraphics[width=.3\textwidth]{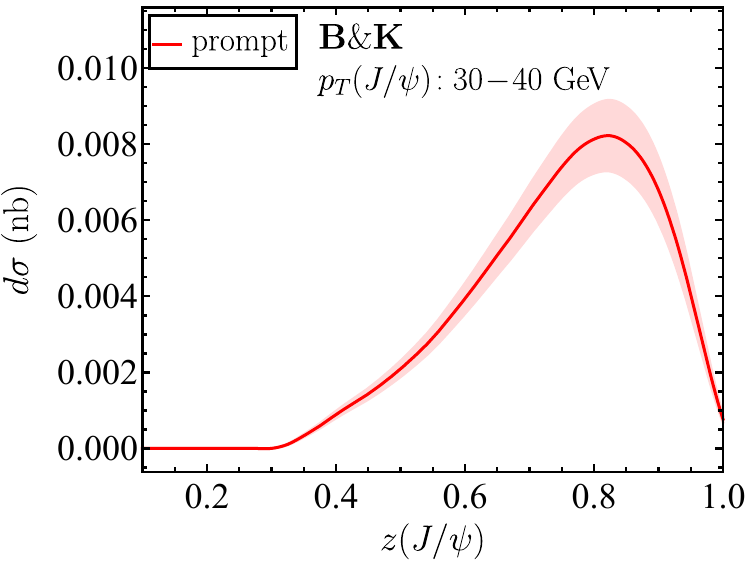}   

	\end{center}\vspace{-4ex}
	\caption{Prediction for $z_{J/\psi}$ distributions for three sets of LDMEs in \tab{ME} and the LHCb measurements.}
	\label{fig:jpsi_LHCb}
\end{figure}
\twocolumngrid

The divergence as $z_H\to 1$ observed in~\cite{Bain:2017wvk} is cured by the threshold resummation, yielding well-behaved predictions in \fig{psi2S_LHCb}. 
However, the LHCb data exhibit pronounced peaks at $z_H=1$ across 
all $p_T$ bins,  except for two highest jet-$p_T$ intervals, 40--60 GeV and 30--40 GeV. These peaks grow more statistically significant at lower $p_T$ values, which may indicate the dominance of a power-suppressed $(m_H/p_T)^2$ double-parton fragmenting process \cite{Fleming:2012wy,Fleming:2013qu,Kang:2014tta,Ma:2013yla,Ma:2014eja,Kang:2014pya}.
For example, in a process $gg \to c\bar{c}g$, where all gluons are coupled to the $c$-quark line, the $c\bar{c}$ pair retains most momentum, producing strong $z_H=1$ peaking.
PYTHIA simulations show similar peaks but broader due to different origins, $g\to c\bar{c}$ transitions, disagreeing with LHCb  data~\cite{LHCb:2024ybz}.

The LHCb prompt-$\psi(2S)$ data demonstrates that binned $z_H$ distributions in both jet-$p_T$ and $\psi(2S)$-$p_T$ significantly improve LDME discrimination and our understanding of the production mechanisms. 
For a comprehensive understanding of LHCb's $z_H$ spectrum, a more quantitative analysis with improved theoretical predictions should be carried out.
High $\psi(2S)$-$p_T$ bins, relatively free from mass corrections and the double-parton process, validate the FJFs framework. 
Low-$z_H$ regions in fixed jet-$p_T$ bins reveal mass correction effects, while the peaks at  $z_H=1$ imply a crucial role of the double-parton process at lower $p_T$ bins. 
This complimentary dual binning makes FJFs ideal to test the NRQCD formalism.
While Ref.~\cite{Bain:2017wvk} identified FJFs' potential as a LDME discriminator through shape differences from the $J/\psi$ data, the $\psi(2S)$ case provides clearer insights. 
The $J/\psi$ analysis is limited by reduced sensitivity to high-$p_T$ region due to  $1/p_T^{5.5}$ scaling of the cross section, which causes the accumulated jet-$p_T$ data dominated by  low-$p_T$ region and less sensitive to high-$p_T$ region.


{\it Prompt-$J/\psi$ results.---} 
\fig{jpsi_LHCb} displays the predicted $z_{J/\psi}$ distributions for the LDME sets listed in \tab{ME} in two regimes: jet-$p_T > 20$ GeV (upper panel) and $J/\psi$-$p_T$ in 30--40 GeV  (lower panel).
Both direct $J/\psi$ production (blue curves) and prompt $J/\psi$ production (red curves) are shown, with the latter including feeddown contributions from excited states.
We note that the B\&K set does not explicitly account for feeddown contributions, as these effects are implicitly absorbed into the LDME values. In the other LDME sets we incorporate feeddown through direct production of  $\psi(2S)$, $\chi_{c1}$, and $\chi_{c2}$ states, which are converted using their respective branching ratios: $\{0.615, 0.195, 0.343\}$.
The LDMEs for S-wave excited states are taken from \tab{ME}, while those for P-wave states adopt Refs.~\cite{Chung:2024jfk,Brambilla:2022ayc,Bodwin:2015iua}. To account for phase-space differences between quarkonium states, their distributions are rescaled via $(m_H/m_{J/\psi})z_{J/\psi}$ as implemented in Ref.~\cite{Bodwin:2015iua}.
The feeddown contribution accounts for approximately 35\% of the total yield in the Brambilla set, while it dominates in the Bodwin set. As observed in the $\psi(2S)$ case, the ${}^3P_J^{[8]} + {}^3S_1^{[8]}$ channels remain the dominant production mechanisms. Notably, the Bodwin set exhibits unphysical negative values in the direct component across the entire $z_H$ range, requiring feeddown contributions to compensate and restore physical positivity in the prompt production yield.

In the upper panel, the predictions from the Bodwin and B\&K sets show reasonable agreement with the LHCb data, while the Brambilla set deviates significantly at low $z_H$. However, this apparent success is tempered by theoretical caveats: the Bodwin set predicts unphysical negative values for direct production, and the B\&K set's treatment of feeddown contributions requires further refinement. Our results remain consistent with lower-order predictions in Refs.~\cite{Bain:2017wvk,Zhang:2024owr} within LDME uncertainties for corresponding LDME sets, despite differences in feeddown treatment.
Unlike $\psi(2S)$ data, the $z_H=1$ peak is not observed, likely due to coarser $z_H$ binning in the region. Higher $z_H$ granularity could resolve this structure. 

The $z_H$ distributions binned in $J/\psi$-$p_T$, analogous to the $\psi(2S)$ analysis, would provide complementary information by suppressing low-$z_H$ subleading corrections and further constraining the production mechanism. Existing $J/\psi$ data from LHCb \cite{LHCb:2017llq} and CMS \cite{CMS:2021puf} could be used to extract these distributions and compare them to our predicted distribution for the 30--40 GeV bin (lower panel of \fig{jpsi_LHCb}). Notably, our predictions were intentionally left unnormalized by area, as both the absolute cross sections and the shape of the distribution offer critical constraints.

\vspace{1cm}

{\it Conclusions.---}
The LHCb $\psi(2S)$-momentum fraction ($z_H$) data, binned in both jet and quarkonium $p_T$, provides an unprecedented level of resolution and enable discrimination between NRQCD LDME sets for $\psi(2S)$.
Our LL+NLO predictions with threshold resummation, when compared with the LHCb data, validate discriminating power of the FJF framework.
Deviations in small $z_H$ highlight the need for mass-suppressed corrections 
and the observed near $z_H=1$ excess suggests double-parton process beyond current leading-power contribution, urging theoretical refinements in the FJF framework.
The $p_T$-integrated $J/\psi$ distributions exhibit moderate agreement for Bodwin (Category 2) and B\&K (Category 3) sets, obscuring clear discrimination, a limitation arising from the coarser $p_T$ resolution.
This analysis establishes quarkonium production within jets as a critical probe of quarkonium-production mechanisms. The dual $p_T$-bin strategy for $\psi(2S)$ suppresses theoretical uncertainties and isolates LDME-dependent features, outperforming integrated $J/\psi$ measurements. Future $J/\psi$ analyses with analogous binning could resolve long-standing LDME tensions. These results underscore the synergy between resolved $p_T$ measurements and effective field theory techniques in unraveling quarkonium production mechanisms.


\begin{acknowledgments}
This work was inspired by Tom Mehen, who initially suggested pursuing this direction. We dedicate the title of this paper to his memory.
D.K.~thanks Korea University for their hospitality during an important phase of this work, 
and acknowledges Chul Kim for fruitful discussions on FJFs. 
Special thanks go to Byungsik Hong, Soohwan Lee, Bayu Putra, and Junseok Lee for valuable discussions regarding the CMS experiment.
The work of D.K. and Y.W. is supported by the National Natural Science Foundation of China (NSFC) through National Key Research and Development Program under the contract No. 2024YFA1610503.
The work of H.~S.~C. is supported by the 
Basic Science Research Program through the National Research Foundation of
Korea (NRF) funded by the Ministry of Education (Grant No. RS-2023-00248313).  
\end{acknowledgments}

\bibliography{FJFbib}

\clearpage

\onecolumngrid

\section{supplementary}

\begin{figure}[thb]
	\begin{center}
		\includegraphics[width=.3\textwidth]{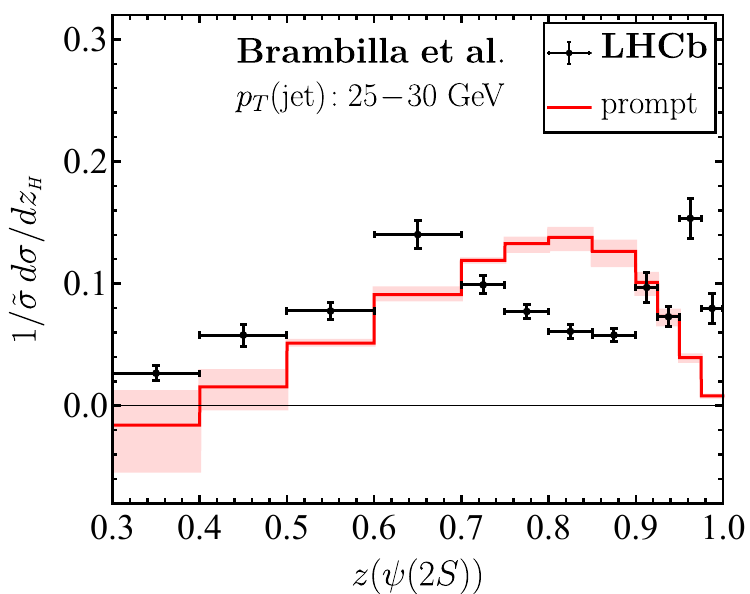}
		\includegraphics[width=.3\textwidth]{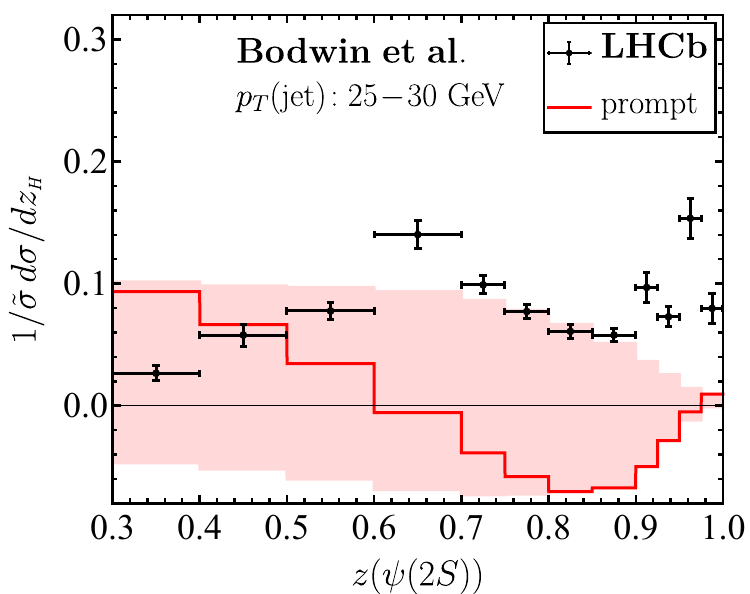}
        \includegraphics[width=.3\textwidth]{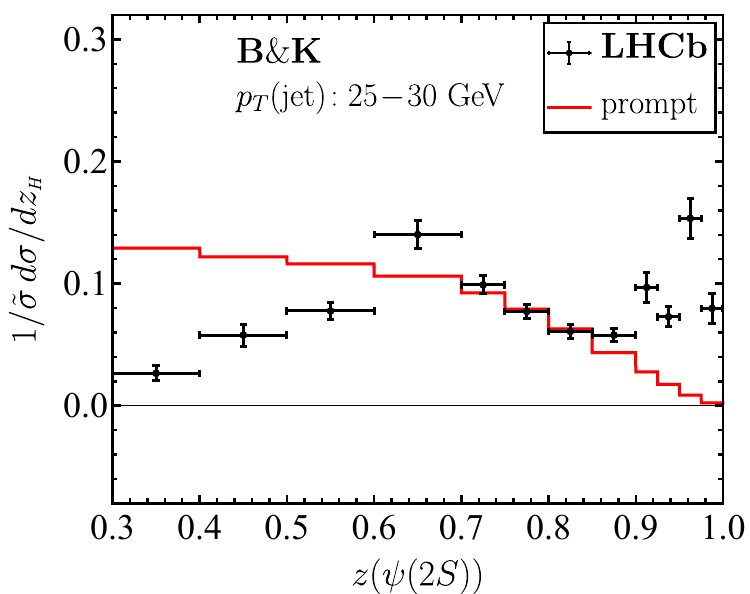}
		\includegraphics[width=.3\textwidth]{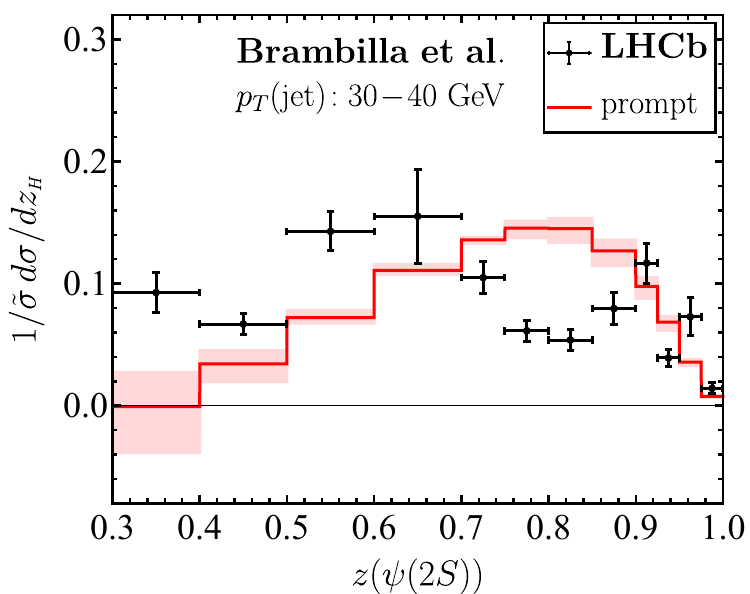}
		\includegraphics[width=.3\textwidth]{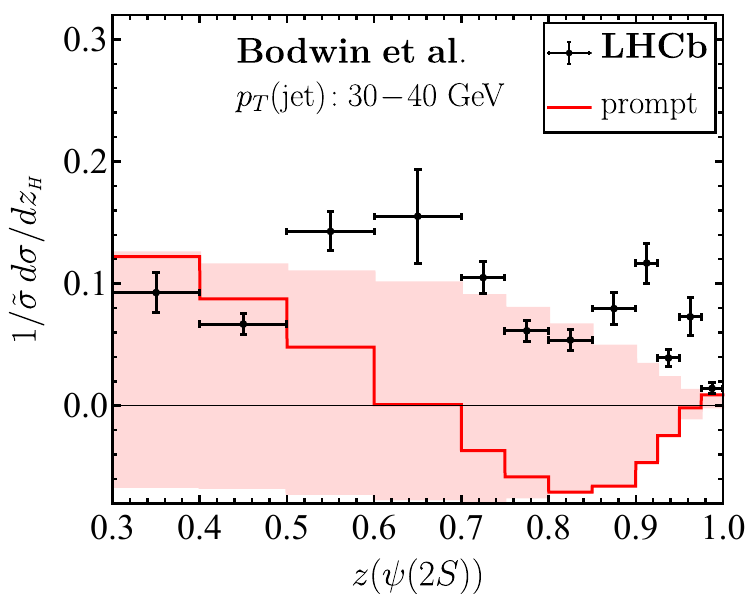}
        \includegraphics[width=.3\textwidth]{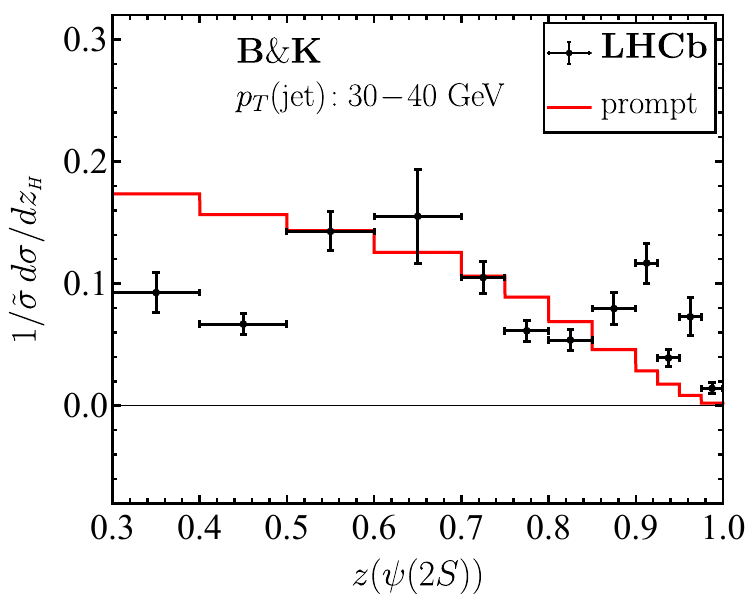}
	\end{center}\vspace{-4ex}
	\caption{Predicted $z(\psi(2S))$ distributions for the three sets of LDMEs  and the LHCb measurements for two jet-$p_T$ bins. }
	\label{fig:psi2S_LHCb2}
\end{figure}

\begin{figure}[thb]
	\begin{center}
		\includegraphics[width=.3\textwidth]{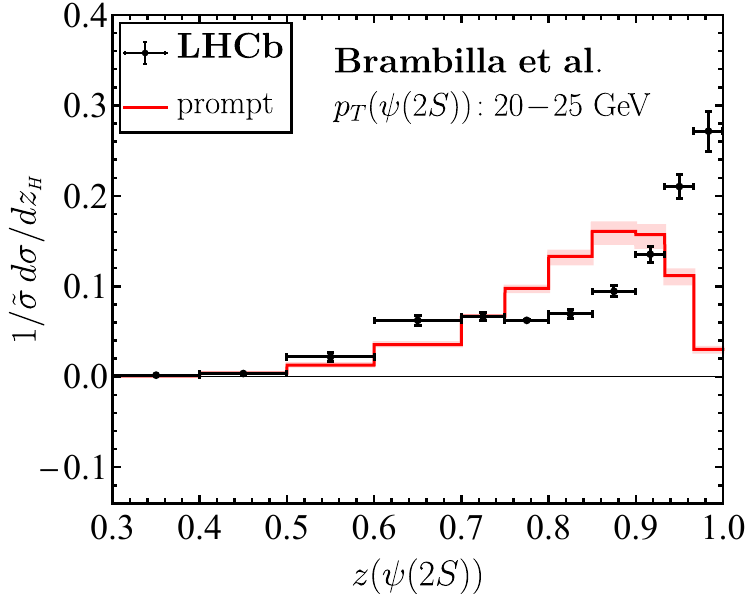}
		\includegraphics[width=.3\textwidth]{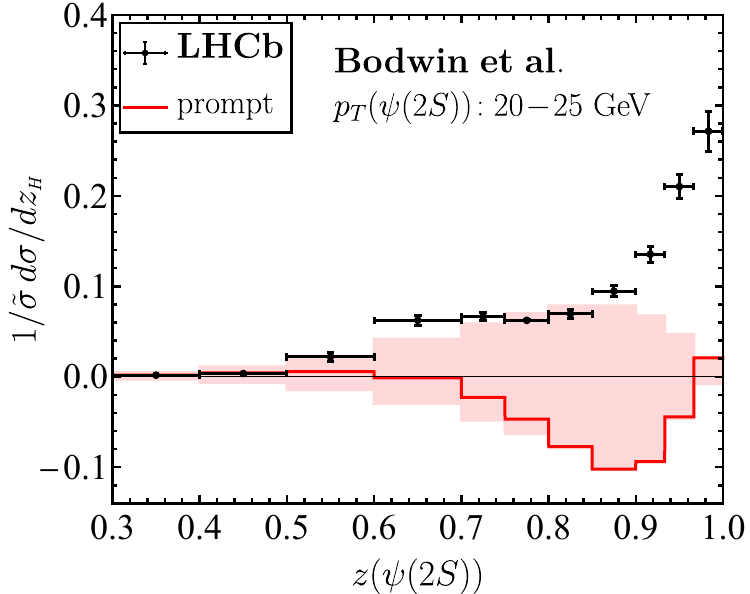}
        \includegraphics[width=.3\textwidth]{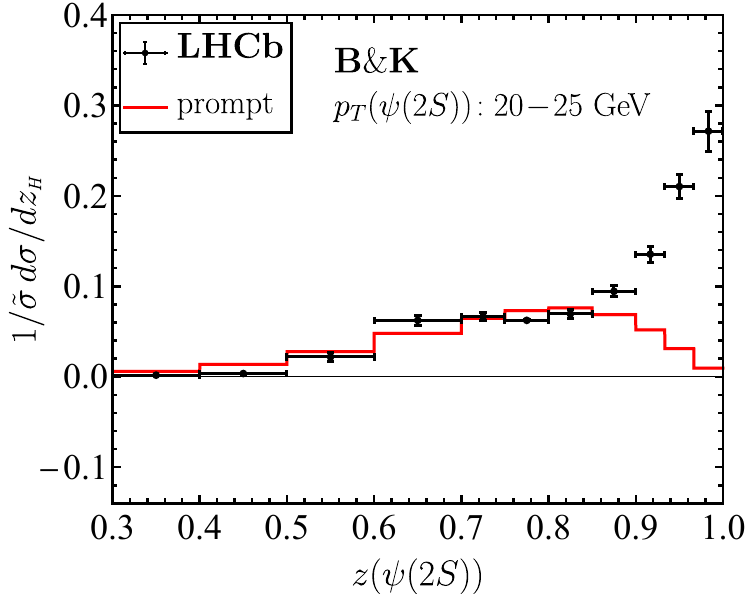}
		\includegraphics[width=.3\textwidth]{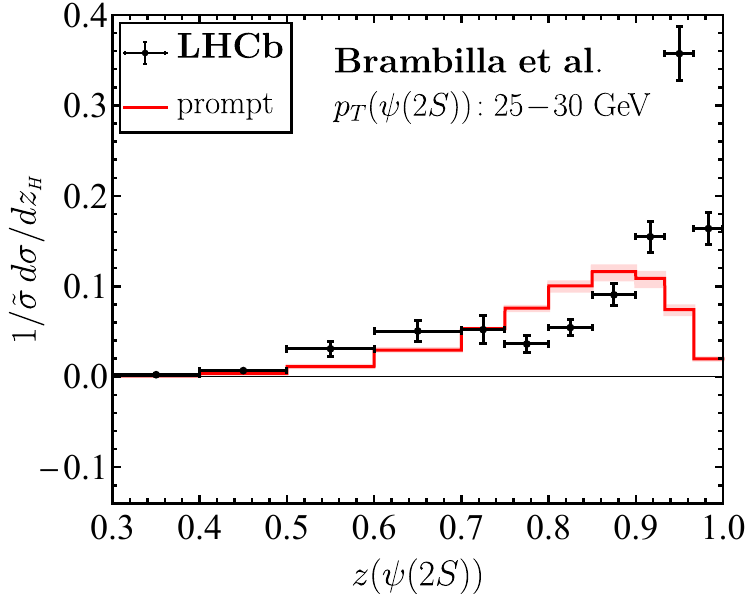}
		\includegraphics[width=.3\textwidth]{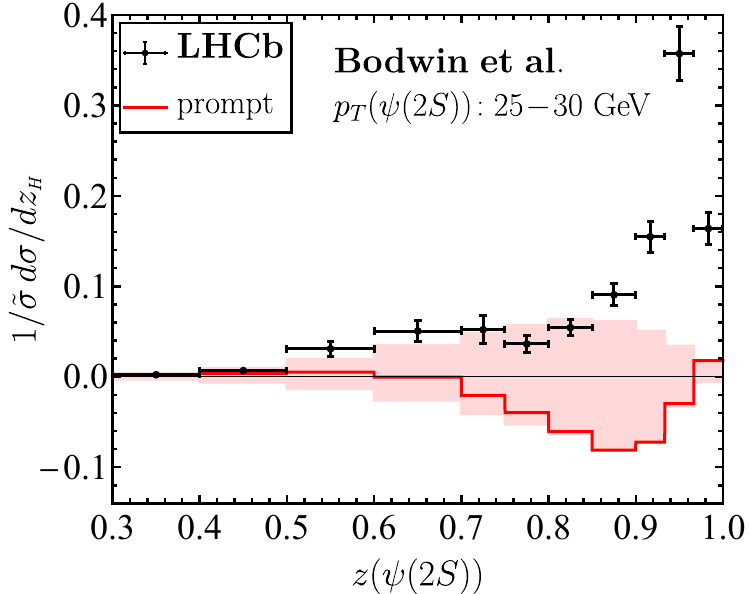}
        \includegraphics[width=.3\textwidth]{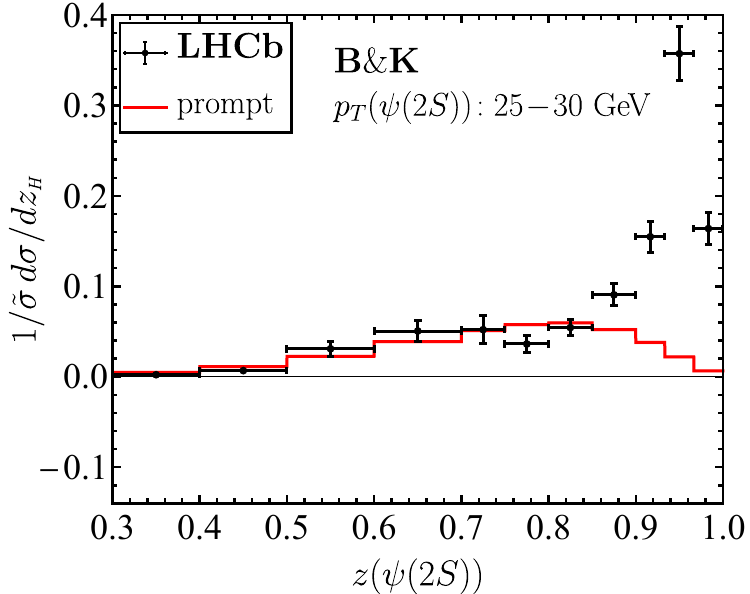}
	\end{center}\vspace{-4ex}
	\caption{Predicted $z(\psi(2S))$ distributions for the three sets of LDMEs  and the LHCb measurements for two $\psi(2S)$-$p_T$ bins. }
	\label{fig:psi2S_LHCb3}
\end{figure}

\twocolumngrid

\end{document}